# The rhombohedral $Sb_2Se_3$ is also an intrinsic topological insulator

G. H. Cao[1], H. J. Liu[1,*], J. H. Liang[1], L. Cheng[1], D. D. Fan[1], Z. Y. Zhang[2,†]

[1] *Key Laboratory of Artificial Micro- and Nano-Structures of Ministry of Education and School of Physics and Technology, Wuhan University, Wuhan 430072, China*

[2] *International Center for Quantum Design of Functional Materials, Hefei National Laboratory for Physical Sciences at the Microscale, and Synergetic Innovation Center of Quantum Information and Quantum Physics, University of Science and Technology of China, Hefei, Anhui 230026, China*

Topological insulators are new class of quantum materials, which have insulating energy gaps in bulk, but exhibit gapless edge states or surface states that are protected by time-reversal symmetry at boundary. It was theoretically predicted and experimentally confirmed that the binary tetradymites $Bi_2Te_3$, $Bi_2Se_3$, and $Sb_2Te_3$ are three-dimensional topological insulators. In this work, we demonstrate by first-principles approach that the ignored $Sb_2Se_3$, although with relatively smaller spin-orbital coupling strength, can also exhibit topologically protected surface states with a bulk gap of 0.19 eV, as long as the van der Waals interaction is explicitly included in the calculations. Detailed analysis of the band structures of $Sb_2Se_3$ thin films indicates that the non-trivial surface state appears at a critical thickness of six quintuple layers.

Topological insulators (TIs) have attracted a lot of attention in material science and condensed matter physics for their fundamental scientific importance and potential applications in spintronics and quantum computing [1, 2, 3, 4, 5]. The study of such new class of quantum materials was primitively inspired by the robustness of gapless edge states in two-dimensional (2D) quantum spin Hall (QSH) systems such as the HgTe/CdTe quantum well, which was theoretically predicted [6] and experimentally confirmed [7] to exhibit metallic edge states protected by time-reversal symmetry

[*] Author to whom correspondence should be addressed. Electronic mail: phlhj@whu.edu.cn
[†] Author to whom correspondence should be addressed. Electronic mail: zhangzy@ustc.edu.cn

(TRS). Subsequently, the QSH effect was generalized to three-dimensional (3D) systems [8]. It was demonstrated, both theoretically and experimentally, that the tetradymites $Bi_2Te_3$, $Bi_2Se_3$, and $Sb_2Te_3$ are all 3D TIs [9, 10, 11]. These binary compounds are stoichiometric crystals with well-defined electronic structures and feature a single gapless Dirac cone in the surface Brillouin zone. However, it seems a bit surprising that the remaining $Sb_2Se_3$ in this family was believed to be a trivial insulator [9, 12], even with the same rhombohedral crystal structure. One may normally think it should be related to relatively smaller spin-orbit coupling (SOC) strength of $Sb_2Se_3$ as compared with the other three tetradymites. This is however not the the real case although the TI nature is usually driven by strong SOC. In this work, we demonstrate by accurate first-principles approach that the ignored $Sb_2Se_3$ is also a topological insulator as long as the van der Waals (vdW) interaction is explicitly considered in the calculations.

Our first-principles calculations have been performed by using the projector augmented-wave (PAW) method [13, 14] within the framework of density functional theory (DFT) [15, 16, 17]. The code is implemented in the so-called Vienna *ab-initio* simulation package (VASP). The exchange-correlation energy is in the form of Perdew-Burke-Ernzerhof (PBE) with the generalized gradient approximation (GGA) [18]. The cutoff energy for the plane-wave basis set is 500 eV. A 15×15×15 Monkhorst-Pack ***k*** mesh [19] is used for the Brillouin zone integrations. For the $Sb_2Se_3$ thin films in few quintuple layers (QLs) regime, a vacuum space of 20 Å is adopted to avoid interactions between the layer and its periodic images. The atomic positions are fully relaxed until the magnitudes of the force acting on each atom become smaller than 0.01 eV/Å. The vdW interactions are explicitly included in our calculations by adopting appropriate functionals [20, 21], which was ignored in previous work [9, 12]. As Sb and Se are both heavy atoms, the SOC effect is taken into account which is also very important for the TI nature.

Figure 1 illustrates the crystal structure of $Sb_2Se_3$, which has a space group of $D_{3d}^5(R\overline{3}m)$. The primitive cell is rhombohedral (Fig. 1(a)) while the unit cell is hexagonal (Fig. 1(b)). The system can be viewed as stacking the QLs along the *c*-direction. Each QL is given in the sequence of Se1-Sb-Se2-Sb-Se1 and they are held together through vdW interactions. During our calculations, the lattice parameters of $Sb_2Se_3$ are optimized by using several different forms of vdW functionals, including the optPBE-vdW, the optB88-vdW, and the optB86b-vdW functionals [20, 21, 22]. In order to check the effectiveness of these vdW functionals, we first calculate the lattice parameters of $Bi_2Se_3$ of the tetradymites family, and compare them with experimentally measured data [23, 24]. As can be seen from Table I, the calculation with optB86b-vdW functional gives lattice constants of *a* = 4.16 Å and *c* = 28.90 Å, which shows the best agreement with the experimental values. In contrast, we find that the standard DFT calculations with PBE functional tend to overestimate the lattice parameters of $Bi_2Se_3$. Note that the optB86b-vdW is also adopted as the best functional in predicting the correct crystal structure of $Bi_2Te_3$ [25]. It is thus reasonable to expect that choosing optB86b-vdW functional could also give a best prediction of the lattice parameters of $Sb_2Se_3$, although the experimental evidence of the rhombohedral structure is still lacking at present. As indicated in Table I, the lattice constant *c* (28.54 Å) is obviously smaller than that predicted by using standard PBE (30.56 Å), and the interlayer distance of 2.61 Å is also smaller than the PBE results of 3.30 Å. All these observations suggest that the vdW interaction is very important and cannot be ignored when dealing with the structural and electronic properties of $Bi_2Te_3$ family and similar layered structures. The vdW functional in the form of optB86b is therefore explicitly used in our following discussions.

The phonon dispersion relation of $Sb_2Se_3$ is shown in Figure 2, which is obtained by adopting the supercell approach with finite displacement method [26]. We see there is no imaginary frequency in the whole phonon spectrum, which suggests that the rhombohedral structure of $Sb_2Se_3$ is kinetically stable. We have also calculated the formation energy of $Sb_2Se_3$ by using the formula $E_{form} = E_{Sb_2Se_3} - 2E_{bulk\_Sb} - 3E_{bulk\_Se}$, where $E_{Sb_2Se_3}$, $E_{bulk\_Sb}$, and $E_{bulk\_Se}$ denotes the total energy of bulk $Sb_2Se_3$, bulk Sb, and bulk Se, respectively. The resulting negative value of −2.9 eV suggests that the rhombohedral Sb$_2$Se$_3$ is also energetically stable. As a further test, we have performed *ab-initio* molecular dynamics (MD) of the system. We choose a microcanonical ensemble and the temperature is set to 300 K. The MD runs for 4600 steps with a time step of 0.5 fs. As shown in Figure 3, there are only small fluctuations around the equilibrium bond lengths of 3.00 Å and 2.82 Å for the nearest Se2-Sb and Se1-Sb distances, and the QL structure of Sb$_2$Se$_3$ remains unchanged even at finite temperature. All these findings suggest that the rhombohedral Sb$_2$Se$_3$ is rather stable.

We now move to the discussions of the electronic properties. Figure 4 plots the orbital-decomposed band structure of the bulk Sb$_2$Se$_3$ along four non-equivalent time-reversal invariant point in the irreducible Brillouin zone. It is clear that the rhombohedral Sb$_2$Se$_3$ exhibit a band gap of 0.19 eV. If we focus on the Γ point, we find that the highest valence band (HVB) and the lowest conduction band (LCB) are mainly occupied by the $p_x$ orbitals of Sb and Se1 atoms, respectively, which obviously exhibit band inversion behavior very similar to those found in the topological insulating Bi$_2$Te$_3$, Bi$_2$Se$_3$, and Sb$_2$Te$_3$ [9, 27]. To confirm that the rhombohedral Sb$_2$Se$_3$ is also a 3D TI, the method proposed by Fu and Kane [28, 29, 30] is adopted to calculate the $Z_2$ invariant, which is based on the analysis of parity for systems with inversion symmetry. Table II summarizes the parity eigenvalues of the occupied bands at eight time-reversal-invariant points. As expected, the $Z_2$ invariant $v_0$ is calculated to be 1, which indicates that the rhombohedral Sb$_2$Se$_3$ is a strong topological insulator [28]. Our calculations thus suggest that all the four binary tetradymites are TIs, and the vdW interaction should be explicitly considered to give correct structural and electronic properties of Sb$_2$Se$_3$.

To check the existence of topologically protected surface states in Sb$_2$Se$_3$, we construct a series thin film with thickness of 1 ~ 9 QLs. It should be emphasized that the optimized structures are also obtained by adopting optB86b-vdW functional as well. In Figure 5, we show the band structures of Sb$_2$Se$_3$ thin films with typical thickness $d$ = 1, 4, 6 and 9 QLs, where the surface states (red circles) are identified by using the wave function projection method [31, 32]. There are two criteria to determine the surface states: in criterion 1, the critical percentages of the projections onto the top-two or the bottom-two *atomic layers* are 45% and 30% for the 1-QL and 2-QL films. In criterion 2, the critical percentages of the projections onto the top-two or the bottom-two QLs are 85%, 70%, 60%, 50%, 45%, 40% and 40% for the 3 ~ 9 QLs films. It should be mentioned that Dai *et al.* [32] proposed that the surface bands are identified by using a critical percentage of the projections onto the topmost or the bottommost QL, while we use the top-two or the bottom-two QLs in criterion 2. Such discrepancy may be attributed to the longer decay length of the surface states of $Sb_2Se_3$. As can be seen from the charge density contour shown in Fig. 5(e), the surface states are primarily localized in the top-two quintuple layers, and decays rapidly at the third QL, which implies the decay length of the surface states of $Sb_2Se_3$ is in the order of 3 nm (or 3 QLs). By counting the times that the surface states cross the Fermi level between time-reversal invariant momenta [30], we can safely confirm that there are only trivial surface states in the 1- and 4-QL films, while there are topologically protected surface states when the film thickness $d \geq 6$ QLs. Moreover, we see from Figure 6 that the band gap decreases monotonically with increasing film thickness and vanishes at a critical thickness of 6 QLs.

Up to now, we have firmly derived that the rhombohedral $Sb_2Se_3$ is a topological insulator. To understand why it is contrary to previous belief [9, 12], we have carried out additional calculations where a 7% tensile strain is applied. The strain is defined as $\varepsilon = (c_s - c_0)/c_0$, where $c_0$ is lattice constant in *c*-direction and $c_s$ is lattice constant under external strain. As discussed above, the interlayer distance we obtained by using optB86b-vdW functional is 2.61 Å. After applying a 7% tensile strain, the interlayer distance becomes 3.22 Å, which is almost the same as those predicted from standard PBE calculations. Figure 7 shows the band structure of $Sb_2Se_3$ under the tensile strain, where we see the previously mentioned band inversion disappears since the HVB and LCB are contributed normally by the $p_x$ orbitals of Se1 and Sb at Γ point, respectively. As a consequence, the $Sb_2Se_3$ becomes a trivial insulator under a tensile strain of 7%, which is just equivalent to previous PBE calculations [9, 12] without considering the vdW effect. Once again, our work emphasizes the very importance of vdW interactions in determining the correct lattice parameters and electronic properties of tetradymites and similar layered structure.

In summary, our accurate first-principles approach demonstrates that only when vdW interactions are explicitly taken into consideration can we give an accurate prediction of the structural and electronic properties of $Sb_2Se_3$, which are usually ignored in previous DFT calculations. In contrast to general belief, we show that $Sb_2Se_3$ is a strong topological insulator with a bulk gap of 0.19 eV, which is verified by calculating the $Z_2$ topological invariants. By using the wave function projection method, the surface states are identified and a phase transition from topologically trivial to non-trivial regime occurs at a critical thickness of six quintuple layers. Although currently there is no direct evidence of the existence of $Sb_2Se_3$ in rhombohedral structure, our theoretical calculations confirm that it is both energetically and kinetically stable, and could be realized with the rapid progress of fabrication techniques. In a word, we conclude that all the binary tetradymites $Sb_2Se_3$, $Bi_2Te_3$, $Bi_2Se_3$ and $Sb_2Te_3$ are topological insulators, even with different SOC strength.

We thank financial support from the National Natural Science Foundation (grant No. 11574236 and 51172167) and the "973 Program" of China (Grant No. 2013CB632502).

**Table I** Structural parameters of Bi$_2$Se$_3$ and Sb$_2$Se$_3$ calculated with various forms of vdW functionals. The previous experimental and theoretical results are also listed for comparison. The unit is angstrom (Å) for the lattice constant and degree (°) for the axial angle.

|  |  |  | hexagonal cell | | rhombohedral cell | |
|---|---|---|---|---|---|---|
|  |  |  | $a$ | $c$ | $a = b = c$ | $\alpha = \beta = \gamma$ |
| Bi$_2$Se$_3$ | References | Expt. 1 (Ref. 23) | 4.138 | 28.64 | | |
|  |  | Expt. 2 (Ref. 24) | 4.143 | 28.636 | | |
|  |  | PBE (Ref. 33) | 4.18 | 30.99 | | |
|  | This work | PBE | 4.19 | 31.26 | 10.70 | 22.57 |
|  |  | optB86b-vdW | 4.16 | 28.90 | 9.93 | 24.20 |
|  |  | optB88-vdW | 4.19 | 29.18 | 10.02 | 24.15 |
|  |  | optPBE-vdW | 4.21 | 29.75 | 10.21 | 23.80 |
| Sb$_2$Se$_3$ | References | PBE (Ref. 9) | 4.076 | 29.830 | | |
|  |  | PBE (Ref. 12) | 4.078 | 29.92 | | |
|  |  | PBE-vdW (Ref. 12) | 4.026 | 28.732 | | |
|  |  | LDA (Ref. 12) | 3.998 | 27.566 | | |
|  | This work | PBE | 4.07 | 30.56 | 10.45 | 22.46 |
|  |  | optB86b-vdW | 4.06 | 28.54 | 9.80 | 23.89 |
|  |  | optB88-vdW | 4.08 | 28.90 | 9.92 | 23.76 |
|  |  | optPBE-vdW | 4.10 | 29.37 | 10.07 | 23.51 |

**Table II** The parity eigenvalues of occupied bands and $Z_2$ invariant $v_0$ of $Sb_2Se_3$. Positive (+1) and negative sign (−1) denote even and odd parity, respectively. $\delta_i$ is the product of parity eigenvalues at each high-symmetry point. The eight time-reversal-invariant points (four non-equivalent time-reversal invariant points) are shown in the inset of Fig. 4.

| band (n) | 1 | 2 | 3 | 4 | 5 | 6 | 7 | 8 | 9 | 10 | 11 | 12 | 13 | 14 | $\delta_i$ | $Z_2(v_0)$ |
|---|---|---|---|---|---|---|---|---|---|---|---|---|---|---|---|---|
| $\xi_{2n}(\Gamma)$ | +1 | 0 | +1 | −1 | +1 | −1 | +1 | +1 | −1 | +1 | −1 | −1 | −1 | +1 | −1 | |
| $\xi_{2n}(Z)$ | −1 | +1 | −1 | +1 | −1 | +1 | +1 | −1 | +1 | +1 | −1 | +1 | +1 | −1 | +1 | 1 |
| $\xi_{2n}(3L)$ | −1 | −1 | +1 | +1 | −1 | +1 | +1 | −1 | +1 | −1 | −1 | +1 | +1 | +1 | +1 | |
| $\xi_{2n}(3F)$ | −1 | +1 | +1 | −1 | +1 | −1 | +1 | −1 | −1 | +1 | −1 | +1 | −1 | −1 | +1 | |

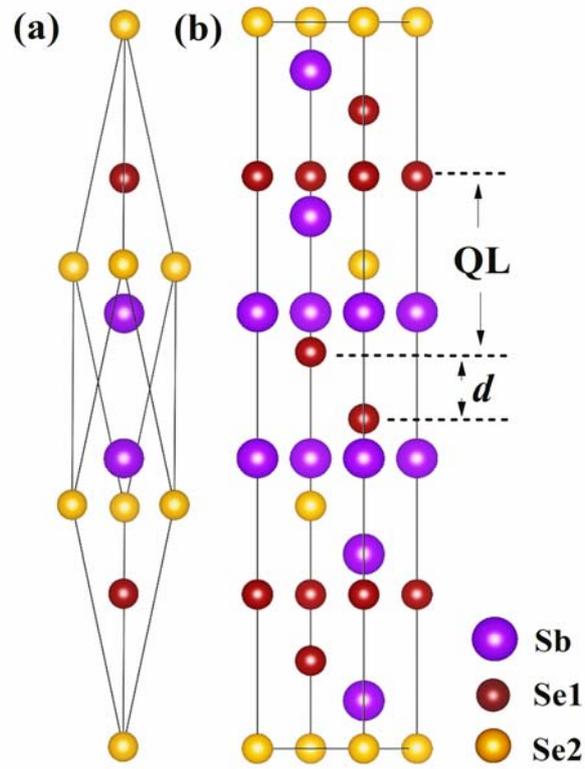

**Figure 1** Optimized crystal structures of the Sb$_2$Se$_3$, where (a) is the primitive cell and (b) is the unit cell. The building block QL and the interlayer distance $d$ are marked.

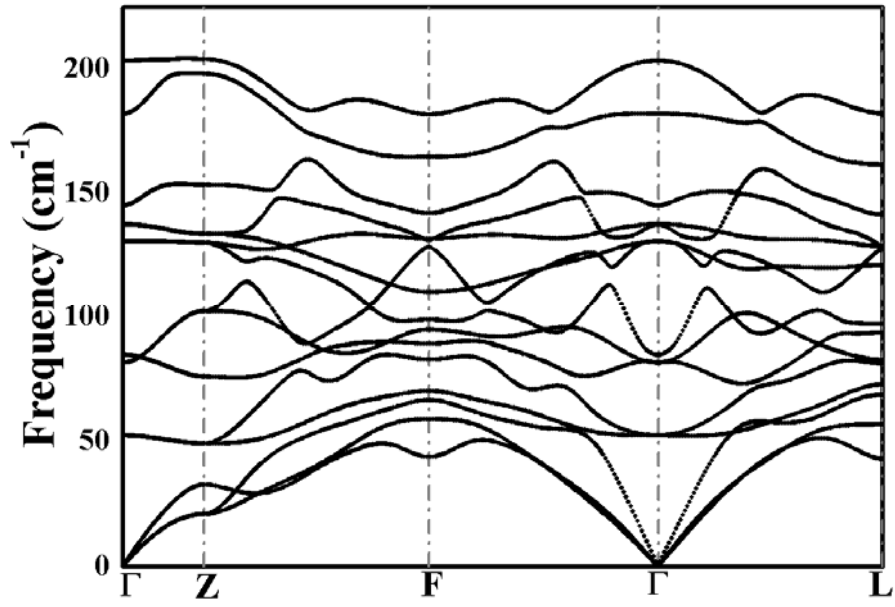

**Figure 2** Phonon dispersion relations of $Sb_2Se_3$.

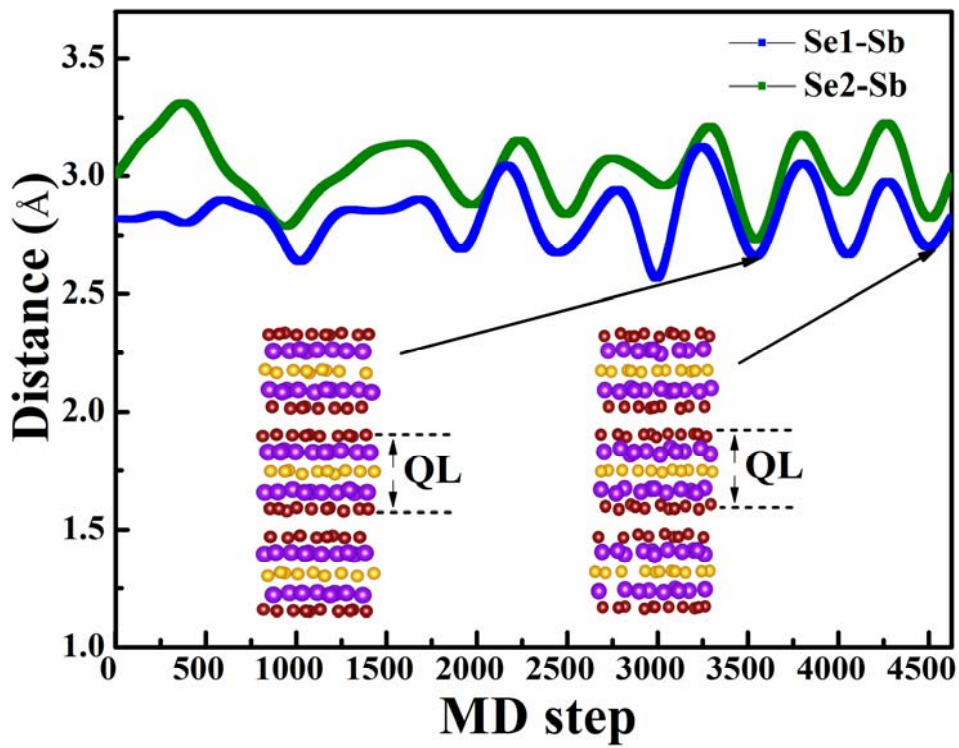

**Figure 3** The *ab-initio* molecular dynamics results of the Se1-Sb and Se2-Sb distances for the Sb$_2$Se$_3$ at room temperature. The two insets correspond the crystal structures at 3546 and 4509 MD step.

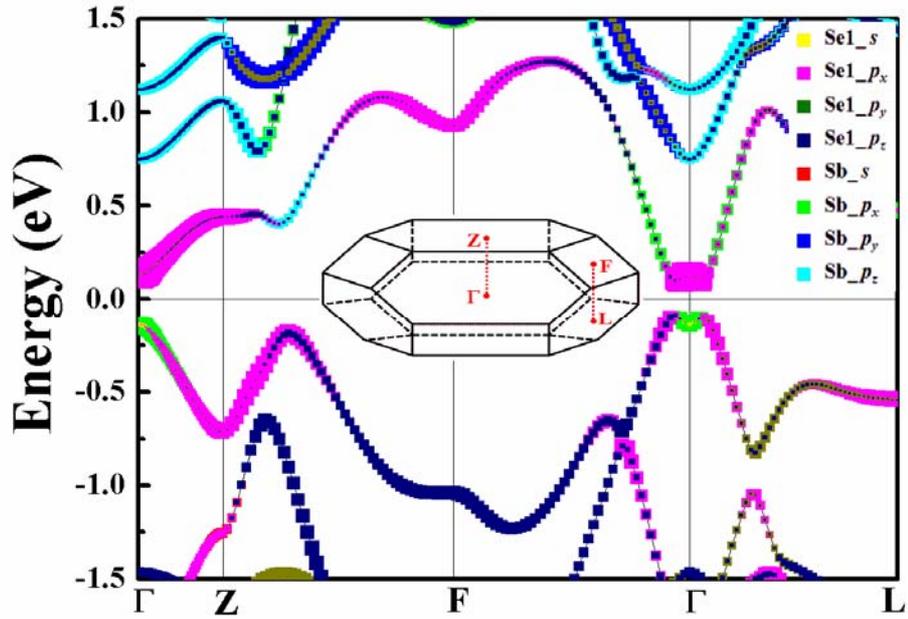

**Figure 4** Band structure of $Sb_2Se_3$ with vdW functional is explicitly taken into account. The inset shows the Brillouin zone of the rhombohedral cell, where four non-equivalent time-reversal-invariant points ($\Gamma$, Z, L, F) are indicated. The colored squares represent the orbital compositions, and the size of the squares correspond the proportions of contribution.

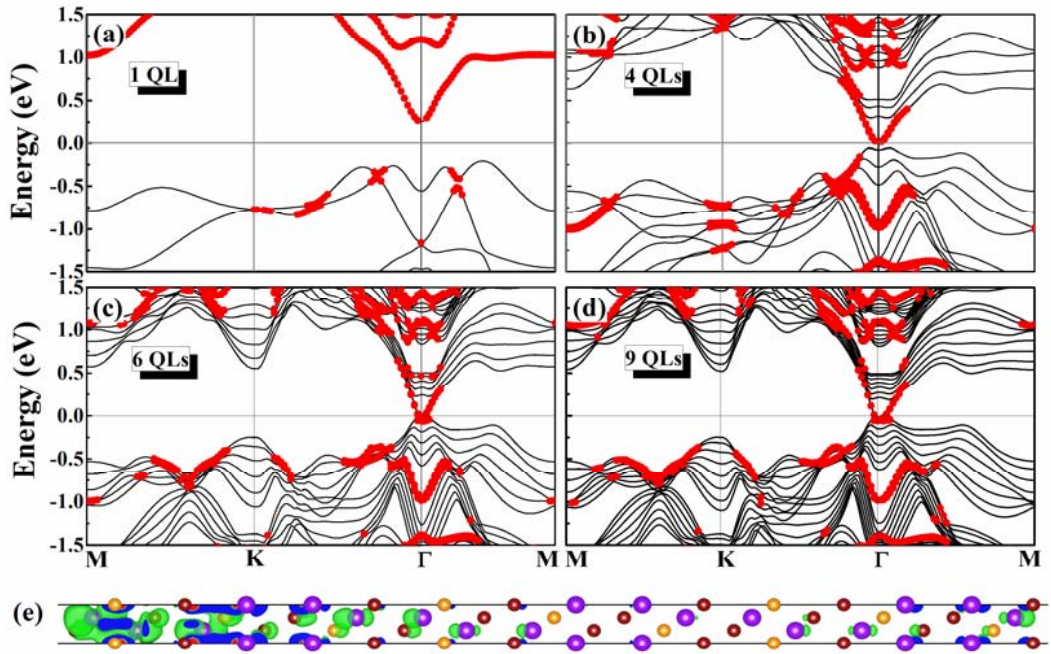

**Figure 5** Band structures of Sb$_2$Se$_3$ thin films with thicknesses $d$ = 1 QL (a), 4 QLs (b), 6 QLs (c), and 9 QLs (d), where the surface states are marked with red circles. The Fermi level is at 0 eV. The charge density contour at Γ point of HVB for the 9-QL film is shown in (e).

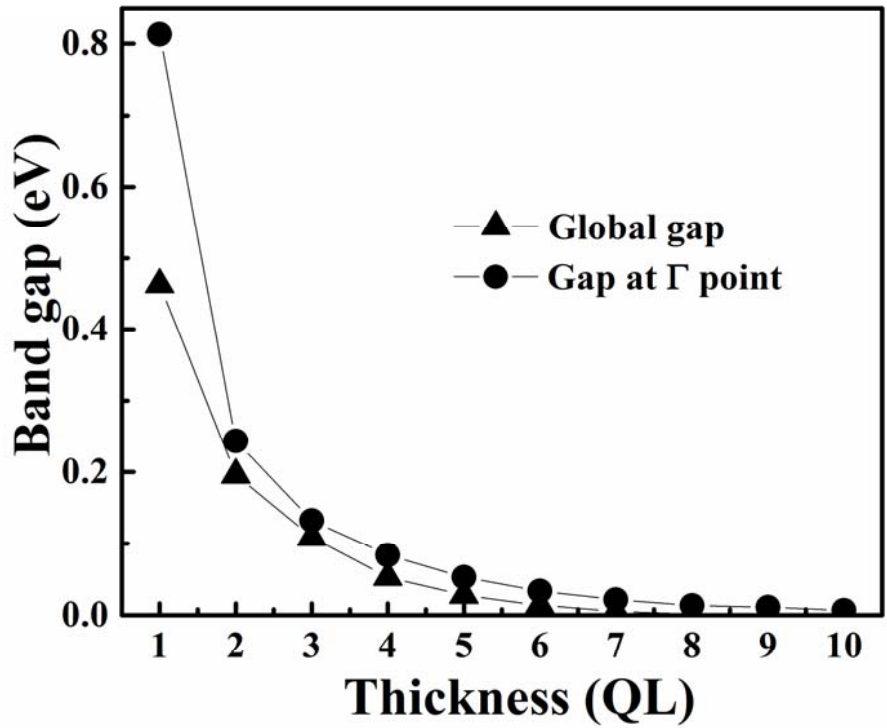

**Figure 6** The band gaps of $Sb_2Se_3$ thin films as a function of thicknesses. The circles and the squares denote the global band gap and the gap at $\Gamma$ point, respectively.

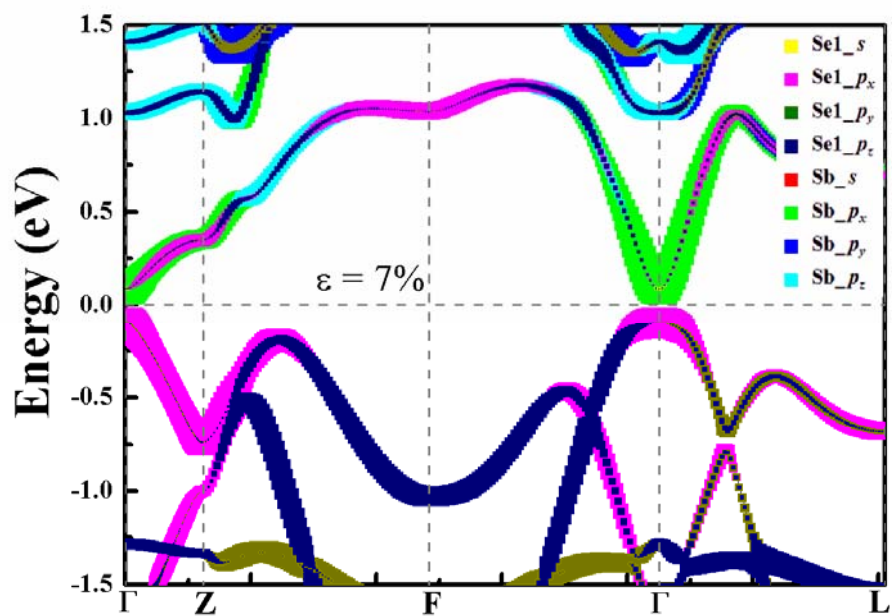

**Figure 7** Band structure of $Sb_2Se_3$ under a 7% tensile strain. The Fermi level is at 0 eV. The color indicates the orbital contribution.